\documentclass[floatfix,twocolumn,aps,prl,showpacs]{revtex4-1}
\usepackage{graphicx}
\usepackage{amsmath}
\usepackage{amssymb}

\newcommand{\inleva}[1]{\langle#1\rangle}
\newcommand{\eva}[1]{\left<#1\right>}
\newcommand{\ket}[1]{\ensuremath{\left| #1 \right\rangle}}

\newcommand{\abs}[1]{\ensuremath{\left| #1 \right|}}

\newcommand{\grad}{\boldsymbol{\nabla}}
\newcommand{\threevec}[3]{\left(\begin{array}{c}#1\\#2\\#3\end{array}\right)}

\newcommand{\wno}[3]{\ensuremath{\left<#1,#2,#3\right>}}
\newcommand{\nematic}{\mathbf{\hat{d}}}
\newcommand{\F}{\mathbf{\hat{F}}}
\newcommand{\absF}{|\langle\mathbf{\hat{F}}\rangle|}

\newcommand{\M}{{\cal M}}
\newcommand{\rr}{\ensuremath{\mathbf{r}}}
\newcommand{\Rtf}{\ensuremath{R_{\rm TF}}}

\newcommand{\phihat}{\ensuremath{\boldsymbol{\hat{\varphi}}}}

\newcommand{\rhohat}{\ensuremath{\boldsymbol{\hat{\rho}}}}

\newcommand{\xhat}{\ensuremath{\mathbf{\hat{x}}}}
\newcommand{\yhat}{\ensuremath{\mathbf{\hat{y}}}}
\newcommand{\zhat}{\ensuremath{\mathbf{\hat{z}}}}
\newcommand{\nhat}{\ensuremath{\mathbf{\hat{n}}}}

\begin{document}

\author{Justin Lovegrove}
\author{Magnus O.\ Borgh}
\author{Janne Ruostekoski}
\affiliation{School of Mathematics, University of Southampton, SO17 1BJ,
    Southampton, UK}

\date{\today}

\title{Energetic stability of coreless vortices in spin-1
  Bose-Einstein condensates with conserved magnetization}

\begin{abstract}
We show that conservation of longitudinal magnetization in
a spinor condensate provides a stabilizing mechanism for a
coreless vortex phase-imprinted on a polar condensate.
The stable vortex can form a composite topological defect with
distinct small- and large-distance topology:
the inner ferromagnetic coreless vortex
continuously deforms toward an outer singular, singly quantized polar vortex.
A similar mechanism can also stabilize a nonsingular nematic texture
in the polar phase.
A weak magnetization is shown to destabilize a coreless vortex in the
ferromagnetic phase.
\end{abstract}
\maketitle

In ordinary superfluids, quantization of circulation around a singular
core is a hallmark of superfluidity.
In superfluids with internal degrees of freedom, circulation need not
be quantized,
and it becomes possible for angular momentum to be carried by
nonsingular textures.
The canonical examples are the Anderson-Toulouse-Chechetkin
(ATC)~\cite{chechetkin_jetp_1976,anderson_prl_1977} and Mermin-Ho
(MH)~\cite{mermin_prl_1976} vortices in
superfluid liquid $^3$He. Coreless spin textures are
also well known in quantum Hall physics~\cite{lee_prl_1990}.

An optically trapped atomic spin-1 Bose-Einstein condensate (BEC)
exhibits two phases of the ground-state manifold.
Coreless vortices
naturally occur in the ferromagnetic (FM) phase, where the order
parameter is determined by spatial
spin rotations and circulation alone is not
quantized~\cite{kawaguchi_physrep_2012,stamper-kurn_arxiv_2012}.
Angular momentum is then carried by a characteristic, nonsingular
fountainlike spin texture~\cite{ho_prl_1998,mizushima_prl_2002,martikainen_pra_2002,reijnders_pra_2004,mueller_pra_2004}.
In the polar phase, by contrast, the expectation value of the spin
vanishes, vortices are singular, and circulation is
quantized~\cite{kawaguchi_physrep_2012,stamper-kurn_arxiv_2012}.

Experiments, however, have demonstrated that the longitudinal
magnetization of the condensate is approximately conserved on the time
scales of current alkali-metal atom
experiments~\cite{stenger_nature_1998,chang_prl_2004,jacob_pra_2012}
and its value can be controlled.
In strongly magnetized BECs, coreless vortices have indeed been prepared by
phase-imprinting in the polar interaction
regime~\cite{leanhardt_prl_2003,choi_njp_2012,choi_prl_2012} where
nonsingular vortices would not be expected to appear by simple
energetic arguments alone.

Here we study coreless, nonsingular vortex textures of spin-1 BECs in
the general case where the system is no longer confined to either the
FM or polar ground-state manifold. We construct analytic models of
their spinor wave functions that interpolate between the two
manifolds. The energetic stability of imprinted coreless
textures is analyzed when the longitudinal magnetization is strictly
constrained throughout the energy-relaxation process. We show that the
coreless vortex, formed by the spin texture, can be energetically
stable in the polar regime for sufficiently strong magnetization. In
the stable configuration, the magnetization leads to a mixing of the
polar and FM phases
with a hierarchical core structure: The interior core region
$\rho=(x^2+y^2)^{1/2}\alt\eta_M$, where $\eta_M$ is the characteristic
length scale determined by the 
magnetization constraint, exhibits a coreless vortex analogous to the
coreless vortex in the FM phase. In the outer region, $\rho\agt
\eta_M$, the coreless vortex continuously deforms toward a singly
quantized, singular polar vortex. 
With distinct small- and large-distance topology, the vortex represents
a \emph{composite topological defect}. Composite vortices also exist
in superfluid liquid 
$^3$He, in which case the hierarchy of different core structures can
result from 
different interaction energies, owing, e.g., to spin-orbit or
magnetic-field coupling~\cite{parts_prl_1995,salomaa_rmp_1987}. 

The vortex core can also be understood in terms of characteristic
energetic length scales. In a spin-1 BEC, where the magnetization is
not conserved, the core size is determined by the larger of the two
healing lengths $\xi_F$ or
$\xi_n$~\cite{ruostekoski_monopole_2003,lovegrove_pra_2012,kobayashi_pra_2012}.
These describe, respectively, the distance over which local
perturbations of the spin expectation value or the density heal. When
the conserved magnetization yields $\eta_M\agt {\rm
  max}(\xi_F,\xi_n)$, it becomes energetically favorable to form a
core of size $\eta_M$, with sharply varying magnetization density.

The phase-imprinting technique employed to create the coreless vortex
with the fountainlike spin texture in
Refs.~\cite{leanhardt_prl_2003,choi_njp_2012} can also be used to prepare a
\emph{nematic coreless vortex}~\cite{choi_njp_2012,choi_prl_2012} in the polar
regime, in which case the characteristic texture is formed by the
nematic axis.  The nematic coreless vortex does not
carry angular momentum, and therefore cannot be stabilized by rotation.
We show that a sufficiently strong magnetization, however, can
energetically stabilize the nematic coreless
vortex. This happens when the magnetization is strong enough to empty one of
the spinor components and form an effective two-component system. The stable
configuration may also be understood in terms
of a composite vortex: in the inner region the nematic axis forms a
fountainlike
coreless texture that continuously deforms in the outer region to a singular,
singly quantized FM vortex. The spin profile then provides a
constraint, stabilizing the nematic coreless vortex inside the core.

Substantial deviations of the conserved magnetization from its `natural' value
can also affect the FM regime. We show that a singly quantized FM vortex can
form the ground state in a rotating condensate.  This may seem
surprising, since
in that case coreless vortices are otherwise predicted
to make up the ground
state~\cite{mizushima_prl_2002,martikainen_pra_2002,reijnders_pra_2004,mueller_pra_2004}.
We find that the spin structure of the coreless vortex cannot support
a \emph{weak}
magnetization and the singly quantized singular vortex, which would
normally exhibit
higher energy than the coreless vortex~\cite{lovegrove_pra_2012},
becomes the new ground state.

We consider the classical mean-field theory of the spin-1 BEC, with
the wave function given by the atom density $n(\rr)$ and a three-component
spinor $\zeta(\rr)$,
\begin{equation}
  \label{eq:wfn}
  \Psi(\rr) = \sqrt{n(\rr)} \zeta(\rr)
            = \sqrt{n(\rr)}
	      \threevec{\zeta_+(\rr)}{\zeta_0(\rr)}{\zeta_-(\rr)},
  \quad \zeta^\dagger\zeta=1\,.
\end{equation}
The $s$-wave interaction between two spin-1 atoms, with
the scattering lengths $a_0$ and $a_2$,
acquires a spin-dependent contribution, yielding two interaction terms
[$c_0=4\pi\hbar^2(2a_2+a_0)/3m$ and
$c_2 = 4\pi\hbar^2(a_2-a_0)/3m$] in the Hamiltonian
density~\cite{supplemental},
\begin{equation}
  \label{eq:hamiltonian-density}
  \mathcal{H} =  \frac{\hbar^2}{2m}\abs{\grad\Psi}^2 +
    \frac{1}{2}m\omega^2r^2n
    + \frac{c_0}{2}n^2
    + \frac{c_2}{2}n^2|\mathbf{\inleva{\hat{F}}}|^2\,,
\end{equation}
where $\omega$ is the trap frequency.
The
spin operator $\F$ is defined as a vector of spin-1 Pauli matrices,
and its expectation value
$\inleva{\F}=\zeta^\dagger_\alpha\F_{\alpha\beta}\zeta_\beta$
yields the local spin vector.

The sign of $c_2$
in Eq.~\eqref{eq:hamiltonian-density} determines which phase is
energetically favored by the
spin-dependent interaction alone. If $c_2>0$, as with $^{23}$Na,
the polar phase with
$\absF=0$ is preferred.  The order parameter is then defined by a
BEC phase $\phi$ and an unoriented nematic
axis $\nematic$~\cite{kawaguchi_physrep_2012,stamper-kurn_arxiv_2012},
such that $\zeta(\phi,\nematic)=\zeta(\phi+\pi,-\nematic)$.
The nematic order leads to the existence of vortices with both integer and half-integer~\cite{leonhardt_jetplett_2000} winding.

Conversely if $c_2<0$, as with $^{87}$Rb,
the FM phase with $\absF=1$ is favored.  All
FM spinors are related by 3D spin
rotations, which define the order parameter space.
This supports two classes of line
defects~\cite{ho_prl_1998,ohmi_jpsj_1998}. The nontrivial
representatives of each class are singular, singly quantized vortices
and nonsingular, coreless vortices, respectively.

The rich phenomenology of the spin-1, and higher-spin, order-parameter
spaces provides
an intriguing system for studies of defects and
textures~\cite{ho_prl_1998,ohmi_jpsj_1998,leonhardt_jetplett_2000,stoof_monopoles_2001, yip_prl_1999,isoshima_pra_2002,mizushima_pra_2002,mizushima_prl_2002,martikainen_pra_2002,ruostekoski_monopole_2003, savage_dirac_2003,zhou_ijmpb_2003,reijnders_pra_2004,mueller_pra_2004,saito_prl_2006, santos_prl_2006,pietila_pra_2007,semenoff_prl_2007,barnett_pra_2007,ji_prl_2008,takahashi_pra_2009, huhtamaki_pra_2009,kobayashi_prl_2009,simula_jpsj_2011,borgh_prl_2012,kobayashi_pra_2012, lovegrove_pra_2012,kawaguchi_physrep_2012,borgh_pra_2013}.
Experimentally, vortices and spin textures have
been investigated by means of both controlled
preparation~\cite{leanhardt_prl_2003,choi_njp_2012,choi_prl_2012,leslie_prl_2009}
and dynamical
formation~\cite{sadler_nature_2006,vengalattore_prl_2008,kronjager_prl_2010,bookjans_prl_2011}.
Here we consider the energetic stability of a coreless
vortex when the magnetization is conserved. The coreless vortex may be
phase-imprinted onto 
a BEC with initially uniform spin texture, regardless of
the sign of $c_2$.
The imprinted vortex state then relaxes due to energy
dissipation and we investigate its structure and
stability.
The magnetization constraint may in general force the wave function
out of its ground-state manifold by mixing the FM and polar phases. 
We will show that the energy relaxation of the prepared vortex can
then be analyzed by introducing the following spinor wave function
\begin{equation}
  \label{eq:cl}
    \zeta^{\rm cl}(\mathbf{r}) =
    \frac{1}{2}
    \threevec{\sqrt{2}\left(f_-\sin^2\frac{\beta}{2}
      - f_+\cos^2\frac{\beta}{2}\right)}
             {-e^{i\varphi}(f_- + f_+)\sin\beta}
             {\sqrt{2}e^{2i\varphi}\left(f_-\cos^2\frac{\beta}{2}
	       -f_+\sin^2\frac{\beta}{2}\right)}\,.
\end{equation}
Here $f_\pm = \sqrt{1 \pm F}$, so that in cylindrical
coordinates $(\rho,\varphi,z)$ the spin vector
$\inleva{\F}=F(\sin\beta\rhohat+\cos\beta\zhat)$, with
$|\inleva{\F}|=F$, displaying the characteristic fountainlike
texture. The coreless vortex \eqref{eq:cl} interpolates between the FM
and polar phases with $0\leq F\leq1$. It can be constructed from a
generalized spinor by combining a $2\pi$ 
winding of the BEC phase with a spin rotation~\cite{supplemental}.

In the limit $F=1$, the spinor \eqref{eq:cl} represents the ordinary
coreless vortex solution in the FM 
phase, with the spin orientation $\beta(\rho)$ increasing
monotonically from zero at $\rho=0$. 
In the opposite limit $F=0$, it becomes a singly quantized polar vortex, with
$\nematic = \cos\beta\rhohat - \sin\beta\zhat$, displaying a radial
disgyration for $\beta\neq\pi/2$~\cite{supplemental}. 
Now we let $F$ vary in space such that $F(\rho\to0)\rightarrow 1$ and
$F\rightarrow 0$ for large $\rho$.  
Equation~\eqref{eq:cl} then describes a \emph{composite} vortex, where the
inner coreless vortex forms the core of the outer polar vortex,
and the circulation $\nu = \oint d\mathbf{r} \cdot \mathbf{v} =
\frac{h}{m}\left(1 - F\cos \beta\right)$ interpolates smoothly between
the corresponding regions~\cite{supplemental}. 

The coreless vortex
is analogous to the nonsingular
ATC~\cite{chechetkin_jetp_1976,anderson_prl_1977} and
MH~\cite{mermin_prl_1976} textures in superfluid liquid
$^3$He~\cite{supplemental}.
For the coreless texture, one
may define a winding number similar to that of a point defect,
\begin{equation}
\label{eq:pi2-charge}
  W= \frac{1}{8\pi} \int_{\cal S} {\rm d}\Omega_i \epsilon_{ijk}
  \nhat_F \cdot
  \left(\frac{\partial\nhat_F}{\partial x_j} \times
  \frac{\partial\nhat_F}{\partial x_k}\right)\,,
\end{equation}
where $\nhat_F=\inleva{\F}/\absF$ and ${\cal S}$ denotes the upper
hemisphere. The charge $W$ counts the number of times $\inleva{\F}$
wraps around the full order parameter space. If the spin vector
reaches a uniform asymptotic 
orientation everywhere away from the vortex (i.e., the bending angle
$\beta(\rho)$ is an integer multiple of $\pi$), $W$ represents an
integer-valued charge. This charge is conserved whenever
the boundary condition is fixed, e.g., by physical interaction or
energetics. If no boundary condition is imposed, the texture can
unwind to the vortex-free state by purely local transformations of the
wave function. The spin-1 coreless vortex may be stabilized by
rotation as the bending angle $\beta(\rho)$ in Eq.~\eqref{eq:cl}, and
therefore the superfluid circulation, adapts to the imposed
rotation. As a result of \eqref{eq:pi2-charge}, coreless textures are
also called 2D `baby Skyrmions'~\cite{manton-sutcliffe} in analogy
with the 3D Skyrmions~\cite{skyrme_1961}, which represent stable
particlelike solitons that can also exist in atomic
BECs~\cite{ruostekoski_prl_2001,battye_prl_2002,savage_prl_2003,ruostekoski_pra_2004,kawakami_prl_2012}.

Magnetic-field
  rotation~\cite{isoshima_pra_2000,leanhardt_prl_2002,shin_prl_2004}
can be used to phase-imprint a coreless
vortex~\cite{leanhardt_prl_2003,choi_njp_2012}.
In Ref.~\cite{choi_njp_2012}, the axial bias field $B_z$
of an external magnetic quadrupole field
$\mathbf{B}
= B^\prime \rho\rhohat + (B_z-2B^\prime z)\zhat$,
is swept linearly such that the zero-field point
passes through an initially spin-polarized BEC $\zeta^1=(1,0,0)^T$~\footnote{In
  Ref.~\cite{leanhardt_prl_2003} a 2D quadrupole field together with an
  axial bias field was used to create a nonsingular vortex. See also
  supplemental material for further details.}.
The sweep amounts to a spin rotation
$\zeta^{\rm i} =
  \exp[-i\F\cdot\beta(\rho)\phihat]\zeta^1$, resulting in the coreless
  vortex~(\ref{eq:cl}), with $F=1$ everywhere. A (pseudo)spin-1 
coreless vortex has also been phase-imprinted in the $\ket{m=0,\pm2}$
levels of a spin-2 $^{87}$Rb BEC (with the $\ket{m=\pm1}$ levels empty)
through population transfer using a Laguerre-Gaussian
laser~\cite{leslie_prl_2009,supplemental}.

By adjusting the spin orientation $\beta(\rho)$ in imprinting
experiments, one can accurately 
control the longitudinal BEC magnetization $M=(N_+-N_-)/N$, where
$N_\pm$ and $N$ are the $\ket{m=\pm1}$ and total atom numbers, respectively.
Owing to the conservation of angular momentum in $s$-wave scattering, the only
spin-flip processes as the energy relaxes are
$2\ket{m=0} \rightleftharpoons \ket{m=+1} + \ket{m=-1}$, which do not
change the magnetization.
Consequently $M$ is approximately conserved as the imprinted
texture relaxes on time scales
where $s$-wave scattering dominates, e.g., over dipolar interactions
and collisions with high-temperature
atoms~\cite{stenger_nature_1998,chang_prl_2004,makela_pra_2007,jacob_pra_2012}.
Dynamical stability of a cylindrically symmetric magnetized coreless
vortex was demonstrated 
against low-energy, cylindrically symmetric Bogoliubov modes in
Ref.~\cite{takahashi_pra_2009}. 

We first study the energetic stability of the phase-imprinted
coreless vortex in the polar regime.
As an initial state we take the magnetically
rotated $\zeta^{\rm i}$ [Eq.~\eqref{eq:cl} with $F=1$ everywhere],
for different $M$.
The energetic stability and structure of the vortex is then determined
by numerically minimizing the free energy
in a rotating trap (at the frequency $\Omega$).
We take
$Nc_0=1000\hbar\omega l^3$ and
$c_0/c_2\simeq 28$ of $^{23}$Na~\cite{knoop_pra_2011} in
Eq.~\eqref{eq:hamiltonian-density},
where $l=(\hbar/m\omega)^{1/2}$.
We strictly conserve the longitudinal magnetization of the initial
state throughout the relaxation procedure. 
The physical mechanism is therefore different from that used in
numerical techniques to achieve a nonzero 
magnetization for the final state by means of an effective linear
Zeeman term, without explicitly conserving the 
magnetization during
relaxation~\cite{yip_prl_1999,isoshima_pra_2002,mizushima_prl_2002,mizushima_pra_2002}. 

In a spin-1 BEC a vortex singularity can be accommodated by exciting the
wave function out of its ground-state manifold, whenever it is energetically
more favorable to adjust the spin value than force the density to vanish at the
singular core~\cite{ruostekoski_monopole_2003,lovegrove_pra_2012}. This
happens
when the characteristic length scales $\xi_n=l(\hbar\omega /2 c_0
n)^{1/2}$ and $\xi_F=l(\hbar \omega/2 |c_2| n)^{1/2}$ satisfy $\xi_F\agt\xi_n$.
Conservation of magnetization
introduces a third length scale $\eta_M$ that describes in the polar
phase the size of a
FM core needed to yield the required $M$.
By considering a uniformly magnetized cylindrical core inside an
otherwise polar density profile, we find by a straightforward integration
the estimate
$\eta_M=\Rtf\sqrt{1-(1-M)^{2/5}}$, where $\Rtf$ is the Thomas-Fermi
radius~\cite{supplemental}.

\begin{figure}[tb]
  \begin{center}
    \includegraphics[width=\columnwidth]{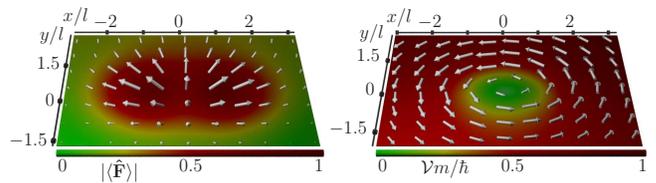}
  \end{center}
  \caption{
    Left: Spin profile $\inleva{\F}$ (arrows, whose length and color
    gradient give $\absF$) 
    of the energetically stable coreless vortex in the polar regime,
    interpolating between 
    FM and polar phases and displaying
    the characteristic fountain texture inside the core of a singular
    polar vortex.
    Right: The corresponding continuously varying superfluid velocity
    $\mathbf{v}$ and its 
    magnitude (arrows),
    and circulation density $\mathcal{V}=\rho\mathbf{v}\cdot\phihat$ (color
    gradient)~\cite{supplemental}.
  }
  \label{fig:polar-cl}
\end{figure}

As the energy of the imprinted coreless vortex relaxes, the outer
region approaches the singly quantized vortex with $F\rightarrow0$.
In the limit of weak magnetization, the vortex core splits into two
half-quantum vortices, similarly to the splitting of a
singly quantized vortex when magnetization is not
conserved~\cite{lovegrove_pra_2012}. 
At $M=0$, the fountain texture is lost entirely. When $M$ increases,
the stable coreless-vortex spin texture gradually becomes more
pronounced, preventing the core splitting. 
The vortex (Fig.~\ref{fig:polar-cl}) still maintains the axial
asymmetry of the magnetized core region with two close-lying 
spin maxima $F=1$, and $\inleva{\F} \parallel \zhat$ at the
center. Ignoring the slight core asymmetry, the vortex can be
qualitatively described by the analytic model~\eqref{eq:cl}: The spin
winds to $\inleva{\F} \parallel \rhohat$ as $\rho$
increases. Simultaneously, $F$ decreases sharply and the configuration
approaches a singly quantized polar vortex. 
The size of the core (the magnetization density half width at half
maximum) is $\sim\eta_M$. 
Comparison of length scales then suggests that the coreless texture
becomes pronounced 
when $\eta_M\agt \xi_F$ ($>\xi_n$), in qualitative agreement with our numerics.

Owing to the trap, $F$ can reach a local minimum---which may not vanish
in all directions---and start increasing at the edge of the
cloud. The vortex profile then depends on $M$ and, e.g., on the
quadratic Zeeman shift that favors the polar
phase~\cite{ruostekoski_pra_2007}. We may therefore 
envisage a scheme to engineer the core symmetry and even more complex
composite defects by Laguerre-Gaussian lasers that generate a Zeeman
shift with a cylindrical shell symmetry. 
\begin{figure}[tb]
  \begin{center}
    \includegraphics[width=\columnwidth]{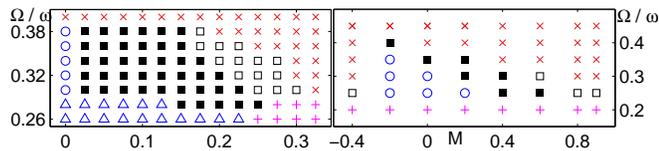}
  \end{center}
  \caption{
    Stability of the coreless vortex in the polar (left) and FM (right)
     interaction regimes;
     ($\blacksquare$) stable coreless vortex;
     ($\square$) stable effective two-component coreless vortex;
    ($\triangle$) instability towards a half-quantum vortex,
     ($\circ$) pair of half-quantum
    vortices (polar regime) or singular vortex (FM regime), ($+$)
    vortex-free state; 
    ($\times$) nucleation of additional vortices;
  }
  \label{fig:stability}
\end{figure}

For stronger magnetization we obtain an effective two-component
coreless-vortex state, where $\zeta_0$ represents a singly quantized
vortex whose core is filled by $\zeta_+$. The transition to the
two-component system occurs when the $\beta(\rho)$ profile no longer
allows the three-component vortex~\eqref{eq:cl} to satisfy the
magnetization constraint, depopulating $\zeta_-$. 
The threshold magnetization value decreases with rotation until at
$\Omega \simeq 0.35\omega$ the coreless vortex is stable only in the
two-component regime (Fig.~\ref{fig:stability}). 
For large $c_0$, the stable two-component solution represents a smooth
transition from the $F=1$ coreless vortex at the trap center to the
$F=0$ singular vortex at the edge of the cloud. 

Next we study the stability of the coreless vortex with conserved magnetization
in the FM regime using $Nc_0=1000\hbar\omega l^3$
and $Nc_2\simeq -5\hbar\omega l^3$ of $^{87}$Rb~\cite{van-Kempen_prl_2002}.
In this case
the FM phase with $F=1$ is preserved everywhere in the imprinted
spinor $\zeta^{\rm i}$
during relaxation. We find that imposed weak magnetization can lead to
energetic
instability of the coreless vortex, with a singular vortex then becoming the
rotating ground state. This may seem surprising, since when the
magnetization is not conserved in energy relaxation, coreless vortices
are 
predicted to form the ground state at sufficiently rapid
rotation~\cite{mizushima_prl_2002,martikainen_pra_2002,reijnders_pra_2004,mueller_pra_2004,lovegrove_pra_2012};
a singular vortex may then also be energetically (meta)stable,
but always has a higher energy than the coreless
vortex~\cite{lovegrove_pra_2012}. 

If magnetization is not conserved, its value for the stable coreless
spin texture at the trap center 
is sensitive to rotation, e.g., varying in the studied case from
$M\simeq0.56$ at 
$\Omega\simeq0.25\omega$ to $M\simeq0.34$ at $\Omega\simeq0.3\omega$.
When the conservation of magnetization imposes weaker magnetization at
a given rotation frequency, we find 
that the vortex is displaced from the trap center as it adjusts to the
constraint. 
At very weak $M$ the coreless vortex is forced close to the edge of
the cloud and splits 
into a pair of singular vortices, one of which exits the BEC.
At faster rotation, the fountain profile bends more sharply,
and consequently
a smaller $M$ can be conserved in the coreless-vortex ground state
(Fig.~\ref{fig:stability})~\footnote{In a strongly magnetized
  condensate, it becomes energetically
favorable to empty $\zeta_-$, stabilizing the coreless vortex
as an effective two-component vortex.}.

The magnetic-field rotation technique used to phase-imprint the FM
coreless vortex~\cite{leanhardt_prl_2003,choi_njp_2012} can also be
applied to a BEC prepared in the polar state $\zeta^0 =
(0,1,0)^T$~\cite{choi_njp_2012,choi_prl_2012}, in which the nematic
axis $\nematic = \zhat$ and $M=0$. The rotation
$\zeta^{\rm p} =
  \exp[-i\F\cdot\beta^\prime(\rho)\phihat]\zeta^0$~\cite{choi_njp_2012}
induced by the magnetic-field sweep then
leads to the nematic texture
$\nematic = \sin\beta^\prime\rhohat + \cos\beta^\prime\zhat$, which defines a
\emph{nematic coreless vortex.} Owing to the magnetic field rotation,
this always
exhibits vanishing longitudinal magnetization.
In Ref.~\cite{choi_prl_2012} an ATC-like (MH-like) texture was imprinted, with
$\beta^\prime(\rho=0)=0$ and
$\beta^\prime=\pi$ ($\beta^\prime=\pi/2$) at the edge of the cloud. The winding
number $W$ may
be defined by taking $\nhat_F \to \nematic$ in Eq.~(\ref{eq:pi2-charge}).
The value of $\nematic$ is not fixed at the boundary,
and the coreless nematic texture may smoothly dissolve. Unlike the
spin texture
of the ordinary coreless vortex, it cannot be stabilized by rotation, due to
its vanishing mass circulation.

We ask instead whether the nematic coreless vortex can be stable
inside the core of
a composite topological defect when a conserved, nonzero magnetization
necessitates
formation of nonpolar regions.
A nematic coreless vortex with a nonzero magnetization could be created by
phase-imprinting via population
transfer~\cite{matthews_prl_1999,andersen_prl_2006,leslie_prl_2009}
that individually prepares 
the appropriate phase windings of $\wno{-2\pi}{0}{+2\pi}$ in the spinor
components.  In order to describe the composite core structure of the
nematic coreless vortex we introduce the spinor wave function
\begin{equation}
  \label{eq:nematic-vortex}
  \zeta^{\rm n}= \frac{1}{2}
    \threevec{\sqrt{2}e^{-i\varphi}\left(f_+\cos^2\frac{\beta}{2}
               -f_-\sin^2\frac{\beta}{2}\right)}
	     {\left(f_+ + f_-\right)\sin\beta}
	     {\sqrt{2}e^{i\varphi}\left(f_+\sin^2\frac{\beta}{2}
	       -f_-\cos^2\frac{\beta}{2}\right)},
\end{equation}
where $\inleva{\F} = F(\sin\beta\rhohat + \cos\beta\zhat)$.
From $\nematic\perp\inleva{\F}$,
it follows that in order to have the fountain texture in $\nematic$ we
must have $\beta=\pi/2$ at $\rho=0$ and increasing monotonically. Then
when $F(\rho)$ increases with $\rho$ from $F=0$ to $F=1$, $\zeta^{\rm
  n}$ interpolates smoothly from the interior
nematic coreless vortex to an outer
singular FM vortex with radial spin component $\inleva{\F}_\rho =
\sin\beta$, with $\beta\simeq\pi$ at the edge of the cloud (MH-like texture).
The corresponding mass circulation,
$\oint d\mathbf{r} \cdot \mathbf{v} =-\frac{hF}{m}\cos\beta$,
interpolates from the noncirculating polar core to a nonzero
circulation, in principle, allowing stabilization by
rotation~\cite{supplemental}.

We find that the numerical relaxation process of the phase-imprinted
$\wno{-2\pi}{0}{+2\pi}$ 
vortex configuration at given magnetization $M$
can be qualitatively described by the spinor~\eqref{eq:nematic-vortex}.
The energetic stabilization of the nematic coreless vortex, however, requires a
negative magnetization $M\alt-0.2$, strong enough that $\zeta_+$ is
depopulated.
The polar phase is then realized only where
$\zeta_0$ fills the vortex line in $\zeta_-$,
with the fountain structure in $\nematic$~\cite{supplemental}.
The core size is again determined by
the magnetization constraint~\cite{supplemental}.   The instability
at weaker magnetization results from the existence of
lower-energy singular vortices with FM cores.

In conclusion, we have shown that the conservation of magnetization in
spinor BECs makes it possible to produce vortex core structures with
distinct small and large-distance topology. The characteristic 
core size is then determined by the magnetization constraint, instead
of one of the healing 
lengths associated with the nonlinear interactions. In
phase-imprinting experiments of 
coreless vortices, the magnetization of the initial vortex state can
be controlled and the 
relaxation process of appropriately prepared initial states results in
mixing of the different 
ground-state manifolds. In particular, a nonsingular, coreless vortex
with sufficiently strong 
magnetization is energetically stable in the polar regime where simple
energetic arguments 
alone would not predict its existence. We have also shown that the
stable configuration can be 
qualitatively understood by an analytic model for the spinor wave
function that interpolates
between the FM and polar phases, representing a composite topological defect.

We acknowledge financial support from the Leverhulme Trust and EPSRC.

\appendix

\setcounter{equation}{0}
\setcounter{figure}{0}
\renewcommand{\theequation}{S\arabic{equation}}
\renewcommand{\thefigure}{S\arabic{figure}}

\section{Supplemental Material}

In this Supplemental online material we provide additional
information regarding spin-1 mean-field theory, conservation of
magnetization, phase imprinting of nonsingular vortices, and
construction of analytic spinor wave functions for the coreless textures.

\section*{Hamiltonian density, magnetization and length scales}

\emph{Hamiltonian density:}
We treat the spin-1 Bose-Einstein condensate (BEC) in the
Gross-Pitaevskii mean-field theory.
The Hamiltonian density then
reads~\cite{kawaguchi_physrep_2012,stamper-kurn_arxiv_2012}
\begin{equation}
  \label{eq:hamiltonian-density-sup}
  \begin{split}
    \mathcal{H} &=  \frac{\hbar^2}{2m}\abs{\grad\Psi}^2 +
    \frac{1}{2}m\omega^2r^2n
    + \frac{c_0}{2}n^2
    + \frac{c_2}{2}n^2|\mathbf{\inleva{\hat{F}}}|^2\\
    &+ g_1n\inleva{\mathbf{B}\cdot\mathbf{\hat{F}}}
    + g_2n\eva{(\mathbf{B}\cdot\mathbf{\hat{F}})^2},
  \end{split}
\end{equation}
where $n=\Psi^\dagger\Psi$ is the atom density and $m$ is the atomic
mass. We have
assumed an isotropic, harmonic trap of frequency $\omega$.
The condensate wave function $\Psi$ is now a three-component spinor,
\begin{equation}
  \label{eq:spinor_suppl}
  \Psi({\bf r}) = \sqrt{n({\bf r})}\zeta({\bf r})
  = \sqrt{n({\bf r})}\threevec{\zeta_+({\bf r})}
                              {\zeta_0({\bf r})}
                              {\zeta_{-}({\bf r})},
  \quad
  \zeta^\dagger\zeta=1,
\end{equation}
in the basis of spin projection onto the $z$ axis.  The local
condensate spin is the expectation value
$\inleva{\F}=\zeta_\alpha^\dagger\mathbf{\hat{F}}_{\alpha\beta}\zeta_\beta$
of the spin operator $\F$, defined as a vector of spin-1 Pauli
matrices.
Linear and quadratic Zeeman shifts, of strength $g_1$ and $g_2$
respectively,
described by the last two terms of Eq.~\eqref{eq:hamiltonian-density-sup}, may
arise from a weak external magnetic field $\mathbf{B}$.

Spin-independent and spin-dependent interaction terms with
strengths $c_0=4\pi\hbar^2(2a_2+a_0)/3m$ and
$c_2=4\pi\hbar^2(a_2-a_0)/3m$, respectively, arise from the two
scattering channels of colliding spin-1 atoms with $s$-wave scattering
lengths $a_0$ and $a_2$. Minimization of the interaction energy then
leads to the two distinct phases of the spin-1 BEC: $c_2<0$ favors the
$\absF=1$ ferromagnetic (FM) phase (e.g., in $^{87}$Rb), while the $\absF=0$
polar phase is favored when $c_2>0$ (e.g., in $^{23}$Na).

\emph{Conservation of magnetization:}
We find stable vortex structures by minimizing the free energy
$E=\int d^3r\,\mathcal{H} - \Omega\inleva{\hat{L}_z}$
in the frame rotating at frequency
$\Omega$ about the $z$ axis, using imaginary-time propagation of the
coupled Gross-Pitaevskii equations.
However, the only spin-flip
processes possible in $s$-wave scattering are
\mbox{$2\ket{m=0} \rightleftharpoons \ket{m=+1} + \ket{m=-1}$.} Therefore
$s$-wave interaction
does not change
the \emph{longitudinal magnetization}
\begin{equation}
  \label{eq:M}
  M = \frac{N_+ - N_-}{N} = \frac{1}{N}\int d^3r\, n(\rr)F_z(\rr)\,,
\end{equation}
where $N$ is the total number of atoms, $N_\pm$ are the populations
of the $\ket{m=\pm1}$ levels and $F_z$ is the $z$ component of the
spin vector.  Consequently, $M$ is approximately
conserved on time scales where $s$-wave scattering
dominates over, e.g., dipolar interactions and collisions with
high-temperature atoms. This is the relevant time scale in present
experiments with spinor BECs of alkali-metal
atoms~\cite{stenger_nature_1998,chang_prl_2004,jacob_pra_2012}.
We take this conservation strictly into account throughout energy
relaxation.

\emph{Characteristic length scales:}
The interaction terms in Eq.~(\ref{eq:hamiltonian-density-sup}) give rise
to the characteristic density and spin healing lengths,
\begin{equation}
  \xi_n = l\left(\frac{\hbar\omega}{2 c_0 n}\right)^{1/2}, \quad
  \xi_F = l\left(\frac{\hbar\omega}{2 |c_2| n}\right)^{1/2},
\end{equation}
where we have introduced the oscillator length
$l=(\hbar/m\omega)^{1/2}$ of the harmonic confinement.
The healing lengths determine, respectively, the length scales
over which the atomic
density $n(\rr)$ and the spin magnitude $\absF$
heal around a local
perturbation.
When magnetization is not conserved, $\xi_n$ and $\xi_F$ determine
the core size of singular
defects~\cite{ruostekoski_monopole_2003,lovegrove_pra_2012}.  If
$\xi_F$ is sufficiently
larger than $\xi_n$, it becomes energetically favorable to avoid
depleting the atomic density, instead accommodating the singularity
by exciting the wave function out of its ground-state manifold. The
core then expands to the order of $\xi_F$, instead of the smaller
$\xi_n$ that determines the size of a core with vanishing density. The
lower gradient energy in the larger core offsets the cost in
interaction energy.

Conservation of magnetization introduces a third length scale
$\eta_M$, which is the size required for a magnetized vortex core
in an otherwise unmagnetized condensate to give rise to a given
magnetization~\eqref{eq:M}.
Here we specifically study a coreless vortex that is phase-imprinted in
the polar BEC. As the energy of the coreless vortex relaxes, the core region
remains magnetized, but the spin magnitude sharply decreases outside the core.
In order to estimate the magnetization length scale
we represent the magnetized core by a cylinder of
radius $\eta_M$, with $\inleva{\F}=\zhat$ everywhere inside, and
$\absF=0$
outside. The total magnetization is then
\begin{equation}
\label{eq:cyl-mag}
  M(\eta_M) = \frac{1}{N}\int d^3r\,
  \Theta(\eta_M-\rho) n_{\rm TF}(\rr)\,,
\end{equation}
where $\rho=(x^2+y^2)^{1/2}$ and $\Theta$ is the Heaviside function.
We approximate the atomic-density profile by the Thomas-Fermi solution
\begin{equation}
  n_{\rm TF}(r)=\frac{15N}{8\pi
  \Rtf^3}\left(1-\frac{r^2}{\Rtf^2}\right),
  \quad r\leq\Rtf\,,
\end{equation}
where $r=(\rho^2+z^2)^{1/2}$, and
\begin{equation}
  \Rtf
    = l\left(\frac{15}{4\pi}\frac{Nc_{\rm p,f}}{\hbar\omega l^3}\right)^{1/5}
\end{equation}
is the
Thomas-Fermi radius. Here $c_{\rm p}=c_0$ in a BEC with polar interactions,
and $c_{\rm f}=c_0+c_2$ in the FM regime.
Computing the integral in Eq.~\eqref{eq:cyl-mag} and solving for
$\eta_M$ as a function of $M$, we obtain
\begin{equation}
  \eta_M = \Rtf\sqrt{1-\left(1-M\right)^{2/5}}\,.
\end{equation}

We also consider a nonsingular coreless vortex formed by a nematic
axis. This was experimentally phase imprinted in a polar BEC in
Refs.~\cite{choi_njp_2012,choi_prl_2012}.
If the imprinted vortex state carries a finite magnetization~\eqref{eq:M},
the relaxation of the structure may be described by a spinor wave
function that interpolates between the polar and FM phases, so that
the core forms a composite topological defect where the interior
represents the polar phase.  We can then define a
length scale $\eta_M^\star$ that describes the size of the
polar core at a given magnetization.  We estimate
$\eta_M^\star$ by taking $\inleva{\F}=\zhat$ everywhere outside a
cylindrical core of this radius, and again
approximating the density profile by the Thomas-Fermi
solution. Then the magnetization is
\begin{equation}
  M(\eta_M^\star) = \frac{1}{N}\int d^3r\,
  \Theta(\rho-\eta_M^\star) n_{\rm TF}(\rr)\,.
\end{equation}
Solving for $\eta_M^\star$ yields
\begin{equation}
  \eta_M^\star = \Rtf\sqrt{1-M^{2/5}}\,.
\end{equation}

\section*{Phase imprinting of nonsingular vortices}

Two different methods have been demonstrated for controlled
preparation of nonsingular vortices. Here we give a brief
overview of each.

In Refs.~\cite{leanhardt_prl_2003,choi_njp_2012} a coreless vortex
was prepared using a time-dependent
magnetic field to induce spin rotations. This technique was first proposed
theoretically in Ref.~\cite{isoshima_pra_2000} and was also
implemented experimentally to prepare singly and
doubly quantized vortices in a spin-polarized
BEC~\cite{leanhardt_prl_2002,shin_prl_2004}.

The creation of a coreless vortex in the spin-1 BEC begins with a
condensate prepared in a fully spin-polarized state, which
we take to be $\zeta^1=(1,0,0)^T$~\footnote{The experiment in
  Ref.~\cite{choi_njp_2012} actually starts from $\zeta=(0,0,1)^T$ and
  creates an ``upside-down'' coreless vortex.}.
The condensate is subject to an external
three-dimensional magnetic quadrupole
field~\cite{choi_njp_2012}
\begin{equation}
  \label{eq:quadrupole}
  \mathbf{B} = B^\prime \rho\rhohat + \left[B_z(t)-2B^\prime z\right]\zhat\,,
\end{equation}
where we have introduced cylindrical coordinates $(\rho,\varphi,z)$.
The zero-field point $z=B_z/2B^\prime$ ($\rho=0$) of the
quadrupole field is initially at large $z$ so that $\mathbf{B}
\parallel \zhat$ in the condensate.

The coreless-vortex structure
is created by linearly sweeping $B_z(t)$ so that the
zero-field point passes through the condensate.  The changing $B_z$
causes the magnetic field away from the $z$ axis to rotate around
$\phihat$ from the $\zhat$ to the $-\zhat$ direction.  The rate of
change of the magnetic field decreases with the distance $\rho$ from
the symmetry axis.  Where the rate of change is sufficiently slow, the
atomic spins adiabatically follow the magnetic field,
corresponding to a complete transfer from $\zeta_+$ to $\zeta_-$ in
the laboratory frame.  However, where the rate of change of the
magnetic field is rapid, atomic spin rotation is no longer adiabatic. In
the laboratory frame, the spins
thus rotate through an angle $\beta(\rho)$, given
by the local adiabaticity of the magnetic-field sweep,
which increases monotonically from zero on the symmetry
axis. Linearly ramping $B_z(t)$ thus directly implements the spin rotation
\begin{equation}
  \label{eq:fm-cl}
  \zeta^{\rm i}(\mathbf{r}) =
  e^{-i\F\cdot\beta(\rho)\phihat}\zeta^1 =
  \frac{1}{\sqrt{2}}\threevec{\sqrt{2}\cos^2\frac{\beta}{2}}
                          {e^{i\varphi}\sin\beta}
                          {\sqrt{2}e^{2i\varphi}\sin^2\frac{\beta}{2}}.
\end{equation}

The resulting fountainlike spin texture
\begin{equation}
  \label{eq:fm-fountain}
  \inleva{\F}=\sin\beta\rhohat + \cos\beta\zhat
\end{equation}
that defines the coreless
vortex in the spinor BEC is analogous to the
Anderson-Toulouse-Chechetkin
(ATC)~\cite{chechetkin_jetp_1976,anderson_prl_1977} and Mermin-Ho
(MH)~\cite{mermin_prl_1976}
vortices that arise in the $A$ phase of superfluid liquid $^3$He.
In liquid $^3$He, a circulation-carrying, nonsingular,
fountainlike
texture is formed by the local angular-momentum vector $\mathbf{l}$ of
the Cooper pairs.  In the ATC texture, $\mathbf{l}$ winds by $\pi$
from the $\zhat$ direction at the center to the $-\zhat$
direction at the edge of the vortex, while the MH texture exhibits a
$\pi/2$ rotation.

The first controlled preparation of a nonsingular
vortex~\cite{leanhardt_prl_2003} used a
two-dimensional quadrupole field together with an axial bias field.
The magnetic field in the trap is then
$\mathbf{B}(\rho,\varphi,\theta) = B_z(t)\zhat +
B^\prime\rho\left[\cos(2\varphi)\rhohat -
\sin(2\varphi)\phihat\right]$.  By the mechanism described above,
ramping of $B_z(t)$ then causes a spin rotation $\zeta(\mathbf{r}) =
  \exp[-i\F\cdot\beta(\rho)\nhat]\zeta^1$ about an axis
$\nhat(\varphi)=\sin\varphi\xhat+\cos\varphi\yhat$. The rotation
yields a
nonsingular spin texture exhibiting a cross disgyration, instead of
the fountainlike structure.  The two are topologically equivalent.

Another technique for phase imprinting a coreless vortex was recently
demonstrated in Ref.~\cite{leslie_prl_2009}.  In this
experiment, the coreless vortex was created in the
$\ket{m=\pm2}$ and
$\ket{m=0}$ magnetic sublevels of the spin-2 manifold of
$^{87}$Rb.  The phase imprinting starts with a spin-polarized condensate in
the $\ket{m=+2}$ level, with a magnetic field along the $z$ axis.
Collinear $\sigma^-$ and $\sigma^+$ polarized laser beams along the
symmetry axis then couple $\ket{m=2}$ to the $\ket{m=0}$ and
$\ket{m=-2}$ levels. The laser beams have Laguerre-Gaussian (LG) and
Gaussian intensity profiles, respectively, so that the population
transferred to the $\ket{m=0}$ ($\ket{m=-2}$) level pick up a $2\pi$
($4\pi$) phase winding. The intensity minimum of the LG beam leaves a
remaining population in $\ket{m=2}$ with no phase winding.  The
resulting five-component spinor represents a coreless vortex with
the spin structure~\eqref{eq:fm-fountain}
when the three nonempty
levels of the five-component spinor are regarded as a (pseudo)spin-1
system.
The bending angle $\beta$ is determined by the density
profiles of the nonempty spinor components.  The laser beams inducing
the Raman coupling of the magnetic sublevels can be tailored with a
high degree of control, and the vortex structure can therefore be
precisely engineered.

By accurately creating specific spin textures, phase
imprinting of coreless vortices
gives control over the longitudinal magnetization
of the cloud, regardless of whether interactions are polar or FM.
In the spin-2 coreless-vortex experiment~\cite{leslie_prl_2009},
the resulting magnetization in the spin-2 manifold is measured at
$M=0.64$ for an imprinted ATC-like spin texture, and at $M=0.72$
for a MH-like texture.
In the magnetic-field rotation experiment~\cite{leanhardt_prl_2003}
the local magnetization
$\M(\rr)=[n_+(\rr)-n_-(\rr)]/n(\rr)$ is reported to be
$\sim 0.7$ at the center of the cloud and
$\sim -0.5$
at the edge.  Because of the lower density in the negatively
magnetized region, also this vortex can be estimated to carry a
positive, nonzero magnetization $M$.

\section*{Generalized vortex solutions}

\subsection*{Coreless vortex}
The two phases of the spin-1 BEC have different order-parameter
symmetries that support different topological defects. For
an overview, see, e.g.,
Refs.~\cite{kawaguchi_physrep_2012,borgh_pra_2013}.  Here we are
interested in nonsingular coreless vortex states that mix the
two phases.  In order to
construct spinor wave functions representing such vortices, we
consider first a representative spinor with uniform spin
$\inleva{\F}=F\zhat$:
\begin{equation}
  \label{eq:zetaF}
  \zeta^F=\frac{1}{\sqrt{2}}\threevec{-\sqrt{1+F}}{0}{\sqrt{1-F}}.
\end{equation}
In particular, the limits $F=0$ and $F=1$ yield
$\zeta^F|_{F=0}=(-1/\sqrt{2},0,1/\sqrt{2})^T$ and
$\zeta^F|_{F=1}=(-1,0,0)^T$,
respectively, corresponding to representative polar and FM spinors.
Any other spinor with $\absF=F$ can be reached by applying a
condensate phase $\phi$ and a three-dimensional spin rotation
$U(\alpha,\beta,\gamma) =
\exp(-i\F_z\alpha)\exp(-i\F_y\beta)\exp(-i\F_z\gamma)$, defined
by three Euler angles, to Eq.~\eqref{eq:zetaF}.  We may thus
write the most general spinor as
\begin{equation}
  \label{eq:general}
  \zeta = \frac{e^{i\phi}}{2}
    \threevec{\sqrt{2}e^{-i\alpha} \left( e^{i\gamma}
      f_-\sin^2\frac{\beta}{2}
      -e^{-i \gamma}f_+\cos^2\frac{\beta}{2}\right)}
     {-\left(e^{i\gamma}f_- + e^{-i \gamma}f_+\right)\sin\beta}
     {\sqrt{2}e^{i \alpha}\left(e^{i\gamma}f_-\cos^2\frac{\beta}{2}
       - e^{-i \gamma}f_+\sin^2\frac{\beta}{2}\right)},
\end{equation}
where $f_\pm=\sqrt{1 \pm F}$.

The state~\eqref{eq:general} can
also be specified by the condensate phase, the spin magnitude $F$ and an
orthonormal triad with one vector in the direction of the spin.
One of the remaining vectors in the triad forms the nematic axis
$\nematic$.  (In the polar limit,
$\nematic$ fully specifies the
state together with the condensate
phase~\cite{ruostekoski_monopole_2003}.) In
the representative spinor~\eqref{eq:zetaF} we choose the triad such
that $\nematic = \xhat$.  In the general spinor~\eqref{eq:general},
$\nematic$ can then be found from the Euler angles.

By allowing $\phi,\alpha,\beta,\gamma$ and the spin
magnitude $F$ to vary in space, we can now construct generalized
solutions representing the coreless vortex states in a condensate
where FM and
polar regions coexist and the wave function interpolates smoothly
between them.
In Eq.~(\ref{eq:general}) we choose $\phi=\alpha=\varphi$ (where
$\varphi$ is the azimuthal angle) and $\gamma=0$ to yield
[Eq.~(3) of the main text]:
\begin{equation}
  \label{eq:cl-sup}
    \zeta^{\rm cl}(\mathbf{r}) =
    \frac{1}{2}
    \threevec{\sqrt{2}\left(f_-\sin^2\frac{\beta}{2}
      - f_+\cos^2\frac{\beta}{2}\right)}
             {-e^{i\varphi}(f_- + f_+)\sin\beta}
             {\sqrt{2}e^{2i\varphi}\left(f_-\cos^2\frac{\beta}{2}
	       -f_+\sin^2\frac{\beta}{2}\right)}.
\end{equation}
The spin texture is then given by
\begin{equation}
  \label{eq:spin-texture}
  \inleva{\F} =
     F(\rr)[\sin\beta(\rr)\rhohat + \cos\beta(\rr)\zhat]\,,
\end{equation}
where $\beta(\rr)$ increases monotonically from zero on the
symmetry axis to form the characteristic fountain texture with
  varying spin magnitude $F(\rr)$.
In the limit $F=1$, we retrieve the coreless vortex
represented by Eqs.~\eqref{eq:fm-cl} and \eqref{eq:fm-fountain}.
In the polar limit $F=0$, on the other hand, Eq.~(\ref{eq:cl-sup})
represents a singly quantized vortex
\begin{equation}
  \label{eq:polar-singular}
  \left.\zeta^{\rm cl}\right|_{F\to0} =
    \frac{e^{i\varphi}}{\sqrt{2}}
    \threevec{-e^{-i\varphi}\cos\beta}
             {-\sqrt{2}\sin\beta}
             {e^{i\varphi}\cos\beta}\,,
\end{equation}
where we have explicitly separated out the condensate phase
$\phi=\varphi$.  The nematic axis forms the texture
$\nematic = \cos\beta\rhohat - \sin\beta\zhat$. In general the spin
rotation that accompanies the winding of the condensate phase
therefore represents a disgyration of the nematic axis.

The vortex \eqref{eq:cl-sup} can represent a solution for which $F$ is
nonuniform, so that 
Eqs.~\eqref{eq:fm-cl} and~\eqref{eq:polar-singular} are the two
limiting solutions. We can form a composite topological defect by
setting $F(\rho=0)=1$ and $\beta(\rho=0)=0$ at the center and letting
$F\rightarrow0$ and $\beta\rightarrow \pi/2$ as $\rho$ increases. Then
the core exhibits a coreless-vortex fountain texture that
continuously transforms toward a singular polar vortex as the radius
increases. 

The mixing of the polar and FM phases in the vortex configuration is
also reflected in the superfluid circulation. 
From Eq.~\eqref{eq:general} we derive the general expression
\begin{equation}
  \label{eq:sf-general}
  \mathbf{v} =
    \frac{\hbar}{m}\grad \phi
    -\frac{\hbar F}{m}\left[\grad\gamma
    + \left(\grad \alpha\right) \cos \beta\right].
\end{equation}
for the superfluid velocity $\mathbf{v}$.
For the coreless vortex~\eqref{eq:cl-sup} this reduces to
\begin{equation}
  \label{eq:cl-v}
  \mathbf{v}^{\rm cl} =
    \frac{\hbar}{m \rho}\left[1 - F(\rho)\cos \beta(\rho)\right]\phihat\,,
\end{equation}
when $F$ and $\beta$ depend only on the radial distance $\rho$.
By considering a circular loop $\mathcal{C}$ at constant $\rho$
enclosing the vortex line,
we can then compute the circulation
\begin{equation}
  \label{eq:circ}
  \nu = \int_\mathcal{C} d\mathbf{r} \cdot \mathbf{v}^{\rm cl}
  =\frac{h}{m}\left[1 - F(\rho)\cos\beta(\rho)\right],
\end{equation}
We may regard the integrand of Eq.~\eqref{eq:circ} as a
\emph{circulation density}
\begin{equation}
  \label{eq:circ-dens}
  \mathcal{V}(\rr) = \mathbf{v}(\rr)\cdot\phihat\rho\,,
\end{equation}
along a cylindrically symmetric path. The circulation of Eq.~(\ref{eq:cl-sup})
continuously interpolates between the polar and FM phases, smoothly
connecting the small-distance 
and large-distance topology of the vortex. Note
that it further follows from Eq.~(\ref{eq:sf-general}) that circulation alone
is quantized only in the limit $F\rightarrow 0$.

Owing to the effectively two-dimensional structure of the coreless
spin texture, it is possible to define a winding number 
\begin{equation}
\label{eq:pi2-charge-sup}
  W= \frac{1}{8\pi} \int_\mathcal{S} {\rm d}\Omega_i \epsilon_{ijk}
  \nhat_F \cdot
  \left(\frac{\partial\nhat_F}{\partial x_j} \times
  \frac{\partial\nhat_F}{\partial x_k}\right)\,.
\end{equation}
Here the integral is
evaluated over the upper hemisphere $\mathcal{S}$, and
$\nhat_F=\inleva{\F}/\absF$ is a unit vector in the direction of the
local spin vector.
The charge $W$ defines a topological invariant if the boundary
condition on $\nhat_F$ away from the vortex is fixed.
When the asymptotic texture is uniform, $W$ is an integer
(representing a mapping between the spin texture and a compactified
two-dimensional plane). 

In the coreless vortex in the spinor BEC,
asymptotic behavior of the spin texture is determined
by rotation, as the bending of $\beta$
in Eq.~\eqref{eq:cl-sup}, and therefore the circulation~\eqref{eq:circ},
adapts to minimize the energy.  The spin texture away from the vortex line
  may also be determined by interactions with other vortices, e.g., in the
formation of a composite defect.
By substituting $\inleva{\F}$ from Eq.~(\ref{eq:spin-texture})
into Eq.~(\ref{eq:pi2-charge-sup}), we may evaluate $W$.  Assuming
cylindrical symmetry and 
taking $R$ to be the radial extent of the
spin texture, we find
\begin{equation}
  \label{eq:cl-W}
  W = \frac{1-\cos\beta(R)}{2}\,,
\end{equation}
where we have used $\beta=0$ on the $z$ axis, such
that $\left.\nhat_F\right|_{\rho=0}=\zhat$. The winding number now
depends on the asymptotic value
of $\beta(\rho)$, such that for $\beta(R)=\pi$ (ATC-like texture)
$W=1$, and for $\beta(R)=\pi/2$ (MH-like texture) $W=1/2$.

\subsection*{Nematic coreless vortex}

From Eq.~(\ref{eq:general}) we can also construct the spinor for a
\emph{nematic coreless vortex} in a magnetized polar BEC.  In
this case we note that
we wish to construct a vortex where
\begin{equation}
  \label{eq:nematic-fountain}
  \nematic = \sin\beta^\prime\rhohat + \cos\beta^\prime\zhat\,,
\end{equation}
corresponding to the state phase-imprinted by Choi et
al.~\cite{choi_prl_2012,choi_njp_2012}. The angle $\beta^\prime$
between $\nematic$ and the $z$ axis increases from $\beta^\prime=0$ at
$\rho=0$ to $\beta^\prime=\pi/2$ ($\beta^\prime=\pi$) at the edge
for a MH-like (ATC-like) texture.
Note that since the Euler angles in Eq.~(\ref{eq:general}) represent
spin rotations of Eq.~(\ref{eq:zetaF}), we have
$\beta=\beta^\prime+\pi/2$, such that $\beta = \pi/2$ at the center of
the vortex.  The desired vortex state can then be constructed by
additionally choosing $\alpha=\varphi$, $\gamma=\pi$ and
$\phi=0$ to yield [Eq.~(5) of the main text]
\begin{equation}
  \label{eq:nematic-vortex-sup}
  \zeta^{\rm n}= \frac{1}{2}
    \threevec{\sqrt{2}e^{-i\varphi}\left(f_+\cos^2\frac{\beta}{2}
               -f_-\sin^2\frac{\beta}{2}\right)}
	     {\left(f_+ + f_-\right)\sin\beta}
	     {\sqrt{2}e^{i\varphi}\left(f_+\sin^2\frac{\beta}{2}
	       -f_-\cos^2\frac{\beta}{2}\right)},
\end{equation}
with spin profile $\inleva{\F}=F(\sin\beta\rhohat+\cos\beta\zhat)$.

In a magnetized BEC, Eq.~\eqref{eq:nematic-vortex-sup} can represent a
composite 
vortex that mixes the FM and polar phases. We consider a solution for which
$F$ exhibits a spatial structure interpolating between $F\rightarrow0$
at the center and $F\rightarrow 1$ 
at the edge of the cloud.  In the limit $F\rightarrow 1$,
Eq.~\eqref{eq:nematic-vortex-sup} becomes 
a singular singly quantized FM vortex,
\begin{equation}
  \left.\zeta^{\rm n}\right|_{F\to1}= \frac{1}{\sqrt{2}}
    \threevec{\sqrt{2}e^{-i\varphi}\cos^2\frac{\beta}{2}}
	     {\sin\beta}
	     {\sqrt{2}e^{i\varphi}\sin^2\frac{\beta}{2}}.
\end{equation}

We calculate the superfluid circulation by assuming $F$ and $\beta$ in
Eq.~(\ref{eq:nematic-vortex-sup}) to be functions of the
radial distance $\rho$ only,
\begin{equation}
\mathbf{v}^{\rm n}= -\frac{\hbar F}{m \rho} \cos \beta\phihat\,.
\end{equation}
Similarly, we find the circulation density
$\mathcal{V}(\rr)=-\hbar F(\rho)\cos\beta(\rho)/m$
and circulation
\begin{equation}
  \label{eq:n-circ}
  \nu = \int_\mathcal{C} d\mathbf{r} \cdot \mathbf{v}^{\rm n}
      = -\frac{hF}{m}\cos\beta\,,
\end{equation}
assuming a circular path $\mathcal{C}$ at constant $\rho$.
It follows that circulation vanishes for the nematic coreless vortex
in the pure polar phase  
$F\rightarrow0$ and becomes nonzero for $F>0$ ($\cos\beta\neq0$). 

\begin{figure}[tb]
  \begin{center}
    \includegraphics[width=\columnwidth]{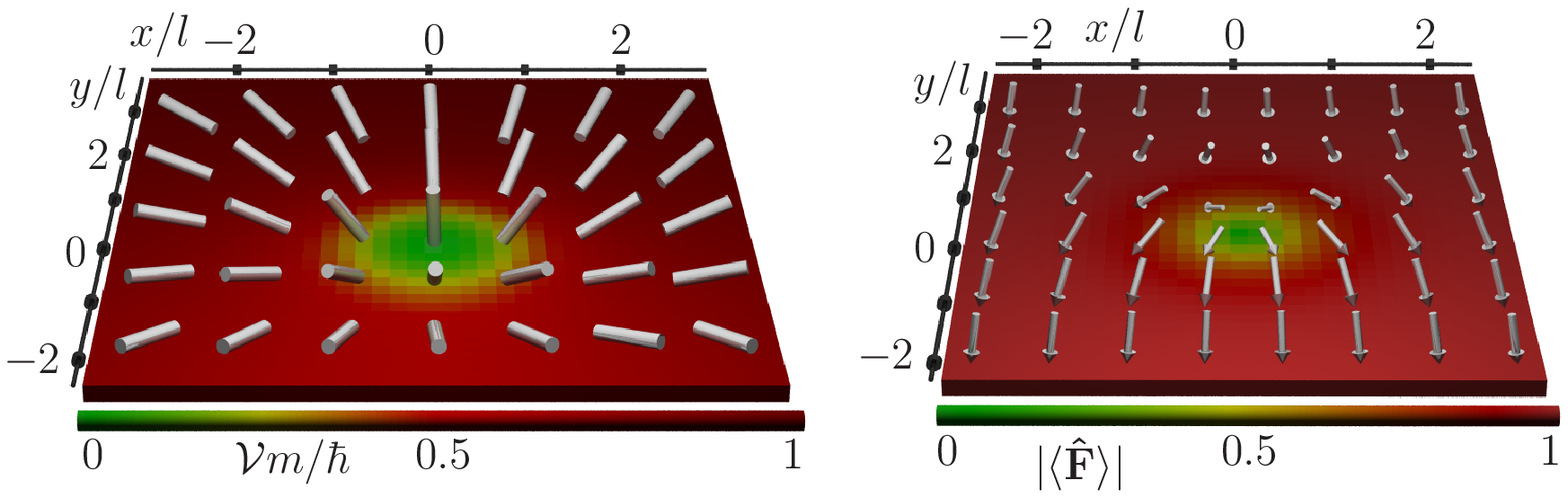}
    \vspace{-0.8cm}
  \end{center}
  \caption{
    Stable coreless nematic
    vortex in a BEC with polar interactions. Left: The unoriented
    $\nematic$-vector
    (cylinders) exhibits the coreless fountainlike texture.  The
    circulation density $\mathcal{V}=\rho\mathbf{v}\cdot\phihat$ (color
    gradient) shows the
    composite-vortex structure that interpolates between the
    noncirculating polar phase to the outer singly quantized FM
    vortex.
    Right: Corresponding spin texture $\inleva{\F}$ (arrows) and spin
    magnitude $\absF$ (color gradient and arrow lengths), showing the
    core region.   Conservation of
    magnetization forces the BEC into the FM phase away from the
    vortex line.
  }
  \label{fig:nematic-cl}
\end{figure}

The interpolation between the polar and FM regimes is illustrated in 
the numerically computed, stable nematic coreless
vortex in both spin magnitude and circulation density in
Fig.~\ref{fig:nematic-cl}. The vortex is energetically stable only
once magnetization is strong 
enough to deplete $\zeta_+$, enforcing an effective two-component
regime.  The stable vortex exhibits a MH-like texture in $\nematic$,
and a corresponding winding 
of the spin vector from the $\rhohat$ direction at the center to the
$-\zhat$ direction in the FM region.

Also for the nematic coreless vortex we may define a
winding number analogous to
Eq.~(\ref{eq:pi2-charge-sup}), associated with the fountain texture of the
nematic axis $\nematic$, by taking $\nhat_F=\nematic$.
Note that due to the equivalence $\nematic
\leftrightarrow -\nematic$ the sign
of $W$ is no longer well defined.
For the cylindrically symmetric fountain
texture~\eqref{eq:nematic-fountain}, the integral in
Eq.~\eqref{eq:pi2-charge-sup} can be evaluated to yield
\begin{equation}
  \label{eq:d-W}
  W = \frac{1-\cos\beta^\prime(R)}{2}
  = \frac{1-\sin\beta(R)}{2}\,,
\end{equation}
making use of $\beta^\prime=0$ on the symmetry axis where
$\nematic=\zhat$. In the last step we have used the relation
$\beta=\beta^\prime+\pi/2$ to rewrite $W$ in terms of the Euler
angle $\beta$ of Eq.~\eqref{eq:nematic-vortex-sup}.
From Eq.~\eqref{eq:d-W}, we find $W=1$ for an
ATC-like texture, and $W=1/2$ for a MH-like
texture such as that stabilized in the effective two-component regime
(Fig.~\ref{fig:nematic-cl}).

\end{document}